\documentclass[aip,twocolumn,letterpaper]{revtex4}

\usepackage{fullpage}

\usepackage{graphics}
\usepackage{epsfig}

\begin{document}
\title{Flattening of the tokamak  current profile by a fast magnetic reconnection with implications for the solar corona}
\author{Allen H. Boozer}
\affiliation{Columbia University, New York, NY  10027\\ ahb17@columbia.edu}

\begin{abstract}

During tokamak disruptions the profile of the net parallel current is observed to flatten on a time scale that is so fast that it must be due to a fast magnetic reconnection.  After a fast magnetic reconnection has broken magnetic surfaces, a single magnetic field line covers an entire volume and not just a magnetic surface.  The current profile, given by $K\equiv\mu_0j_{||}/B$, relaxes to a constant within that volume by Alfv\'en waves propagating along the chaotic magnetic field lines.  The time scale for this relaxation determines the commonly observed disruption phenomena of a current spike and a sudden drop in the plasma internal inductance.  An efficient method for studying this relaxation is derived, which allows a better understanding of the information encoded in the current spike and the associated sudden drop in the plasma internal inductance.  Implications for coronal heating are also discussed.

\end{abstract}

\date{\today} 
\maketitle


\section{Introduction}

During disruptions on the largest operating tokamak, JET, the current profile, $K\equiv\mu_0j_{||}/B$, flattens on a time scale $\lesssim 1$~ms, which is orders of magnitude shorter than the time scale for resistive diffusion.  The evidence  \cite{Wesson:1990,de Vries:2016} for the flattening is a current spike,  which is an increase in the net plasma current, and a drop in the plasma internal inductance, which is a measure of radial width of $K$.

The flattening of $K$ on a time scale orders of magnitude shorter than possible with resistive diffusion was found to require magnetic surface breakup and magnetic helicity conservation in a 1991 numerical study by Merrill, Jardin, Ulrickson, and Bell \cite{Jardin:1991}.  The speed of surface breakup can be understood as a fast magnetic reconnection \cite{Boozer:ideal-ev}.  A fast magnetic reconnection breaks magnetic surfaces on time scale primarily determined by the properties of the evolution, not resistivity, and flattens the current profile on a time scale determined by the Alfv\'en speed.  The spike in the current is an implication of  a helicity-conserving current flattening on a time scale short compared to the resistive time scale.  The physics of the flattening through shear Alfv\'en waves is the focus of this paper.   

The breakup of magnetic surfaces is of central importance to understanding the danger of runaway electrons to ITER \cite{Boozer-spikes:2016}.  When magnetic field lines from a large fraction of the plasma volume can reach the walls, the energetic electrons that serve as a seed for electron runaway are quickly lost.  The magnitude of the current spike is a measure of the volume in which magnetic surfaces have been destroyed \cite{Boozer-spikes:2016}.  However, a thermal quench, a large drop in the electron temperature, preceeds or accompanies the current spike, and the enhanced dissipation of magnetic helicity associated with the lower temperature reduces the magnitude of the spike and complicates its interpretation.  

An equation derived using a mean-magnetic-field approximation \cite{Boozer:surf-loss} could be used to study the spatial and temporal extent of the breakup of magnetic surfaces when reliable measurements of the plasma current, internal inductance, and the electron temperature are available. This analysis would be simpler and faster than that in \cite{Jardin:1991}.  The mean-field approximation does not address the Alfv\'enic process that gives the flattening of $K$, which is the focus of this paper, but uses the helicity-conserving property of a fast magnetic reconnection \cite{Boozer:acc} to obtain an differential equation of the simplest physically-consistent form for the evolution of $K$.

Section \ref{sec:background} gives background information on three topics: (1) fast magnetic reconnection, Section \ref{sec:fast}, (2) the phenomenology of tokamak disruptions, Section \ref{sec:phenomenology}, and (3) the drive and damping of Alfv\'en waves, Section \ref{sec:Alfven}.  Those familiar with this material can go directly to Section \ref{sec:flattening}, which derives the equation for the Alfv\'en waves that flatten the current.   Section \ref{sec:Monte-Carlo} explains how this equation can be solved for the current flattening using a Monte-Carlo method.  Section \ref{sec:discussion} discusses the paper and its conclusions. 


\section{Background information \label{sec:background} }

\subsection{Fast magnetic reconnection \label{sec:fast} }

Fast reconnection arises naturally when an evolving magnetic field depends on all three spatial coordinates \cite{Boozer:ideal-ev}.   The magnetic field line velocity $\vec{u}$ of an ideal evolution can exponentially distort magnetic surfaces or more generally magnetic flux tubes.  This distortion leads to a multiplication of the non-ideal effects by a factor that increases exponentially on a time scale determined by the ideal evolution.  Large current densities are not required for a fast reconnection. 

Magnetic flux tubes are an essential concept for understanding magnetic fields that are smooth functions of the spatial coordinates.  There need be no implication that the field is unusually strong in the interior of a flux tube as is often the case in the astrophysical literature.   The surface of a flux tube is formed by the field lines that pass through a particular closed curve---often taken to be a circle of radius $r_c(0)$.   The cross-sectional shape of a flux tube distorts with distance $\ell$ along the tube.  As $r_c(0)\rightarrow0$, the distortion becomes simple, an ellipse.  Since $\vec{\nabla}\cdot\vec{B}=0$, the  major $r_\ell(\ell)$ and minor $r_s(\ell)$ radii of the ellipse satisfy $ r_\ell r_s=r_c^2$ when $r_c(\ell)$ is defined so $B(\ell)r_c^2(\ell)$ is constant.  The exponentiation $\sigma_e(\ell)$ is defined by $r_\ell=r_c e^\sigma_e$ and $r_s=r_c e^{-\sigma_e}$.  For all but special ideal field line flows $\vec{u}$, the cross-sectional distortion of a given flux tube becomes larger as time advances \cite{Boozer:ideal-ev}; typically $\sigma_e$ is approximately proportional to time.   

Resistive magnetic-reconnection competes with the ideal evolution when the time required for resistive diffusion over the small distance across a flux tube $r_s^2\mu_0/\eta$ competes with the evolution time scale, $\tau_{ev}\equiv 1/|\vec{\nabla}\vec{u}| \approx r_c/u$, where $r_c$ is a characteristic initial dimension of the tube.  The resistive time scale is defined by $\tau_\eta\equiv r_c^2\mu_0/\eta$, so resistive diffusion competes with the ideal evolution when $R_m\equiv \tau_\eta/\tau_{ev} = e^{2\sigma_e}$, and a fast magnetic reconnection occurs.  $R_m$, the magnetic Reynolds number, is of order $10^4$ to $10^8$ in tokamaks and up to $10^{14}$ in the solar corona.  The required current density to produce an exponentiation $\sigma_e$ is proportional to $\sigma_e$ or equivalently to $\ln(\sqrt{R_m})$ and not $R_m$, as required for reconnection to compete with evolution without exponentiation.

 In a two-dimensional ideal evolution, exponentially large distortions of flux tubes require an exponentially large change in the magnetic field strength, but no change in the magnetic field strength is required for exponentially large distortions in a three-dimensional ideal evolution \cite{Boozer:ideal-ev}.  
 
 Magnetic reconnection can be studied in three dimensions ignoring the effect of exponentiation, but this is as misleading as ignoring the non-diffusive advective stirring of air when calculating the time it takes a hot radiator to warm a room.  The ideal advection of air is a divergence-free flow, which causes tubes of air-flow to distort exponentially, just as magnetic flux tubes distort, which enhances diffusive mixing.  Exponentiation changes the time scale for warming a room  from several weeks to tens of minutes.  In three-dimensional simulations of reconnection, one can verify that reconnection occurs where the exponentiation is large, as was done by Daughton et al  \cite{Daughton:2014}.  Numerical resolution limited the number of exponentiations that they could observe to $\sigma_e\lesssim8$, which implies their code can resolve the physics only when $R_m \lesssim 10^7$.

A fast magnetic reconnection conserves magnetic helicity \cite{Boozer:acc} with even greater accuracy than the limit given in 1984 by Berger \cite{Berger:1984}.  Helicity conservation requires an increase in the plasma current when the current profile, $K=\mu_0j_{||}/B$, is flattened \cite{Boozer-spikes:2016}.  

As will be discussed, the time scale for the flattening is determined by the time required for a shear Alfv\'en wave to propagate along magnetic field lines.   To obtain a current spike on the observed sub-millisecond time scale, chaotic magnetic field lines must cross a large fraction of the $j_{||}/B$ profile and reach the edge of the plasma in of order a hundred toroidal transits.  In JET, a shear Alfv\'en wave requires $\approx 3~\mu$s to make a full toroidal transit. A hundred transits is comparable to the independent observations in a numerical simulation of a tokamak disruption by Valerie Izzo \cite{Izzo:2020} and that by Eric Nardon et al., which is not yet published.

The speed of the flattening rules out simple resistive diffusion as an explanation.  The time scale for resistive diffusion of the current density using a cylindrical model is $\tau_j = (\mu_0/\eta)(a/2.40)^2$.   The resistive diffusion coefficient, $\eta/\mu_0\approx 2\times10^{-2}/T^{3/2}$ where the temperature is in keV, distances in meters and times in seconds.  Plasma cooling precedes or accompanies current flattening.  But, even at the lowest estimated plasma temperature of 10~eV,  $\eta/\mu_0\approx 20$, and $\tau_j\approx9~$ms in JET, which has a minor radius $a\simeq1~$m.  The flattening takes place on a sub-millisecond time scale.

In the solar corona, it is the motion of the magnetic field lines on the photosphere that is thought to drive what is initially an ideal evolution, which ultimately leads to a fast magnetic reconnection.   In tokamak disruptions, the ideal drive is an increasingly contorted annulus of magnetic surfaces between low order magnetic islands.  These islands grow at a rate that appears to be consistent with the Rutherford rate \cite{Rutherford:rate}.  As illustrated in de Vries et al \cite{de Vries:2016}, JET shows a sudden acceleration in the evolution from a Rutherford-like slow growth of non-axisymmetric magnetic fields to a current spike and a drop in the internal inductance that evolve approximately three orders of magnitude faster, which is also much faster than the time scale of the observed subsequent current quench, $\sim20~$ms, which occurs after the thermal quench.


\subsection{Phenomenology of tokamak disruptions \label{sec:phenomenology} }

Tokamak disruptions have many causes.  Sometimes they are initiated by intentional impurity injection, which produces strong radiative cooling early in the disruption and can result in strong currents of relativistic electrons that allow studies of the behavior of such currents in tokamaks.  A more serious issue for ITER is naturally arising disruptions, but details of only a few examples have been published.  Two examples have been published for JET, but even these lack important details.   

Figure 1 in the classic paper by Wesson et al \cite{Wesson:1990} on a carbon-wall JET disruption showed a drop in the central electron temperature from 1.6~keV to  0.5~keV starting approximately 3.5~ms before the current spike.  This temperature drop was supposedly due to internal MHD activity breaking the magnetic surfaces and flattening the temperature  in the inner half of the plasma.    Resistive diffusion would require approximately 10~s, so a fast magnetic reconnection is required.  As Wesson et al noted, intact outer magnetic surfaces would shield the outer world from a current spike, and even at 10~eV the breakup time for the outer magnetic surfaces would be of order 10~ms.  The current spike occurs over 200~$\mu$s, and the electron temperature drops from 500~eV to an estimated 10~eV within 300~$\mu$s.  This temperature drop was thought to be due to an impurity influx.  As discussed below, heat flow along chaotic magnetic field lines could easily explain the drop in temperature from 1.6~keV to  0.5~keV, but heat flow along chaotic field lines becomes extremely slow at lower temperatures and impurity radiation seems the only credible explanation for reaching 10~eV.  The subsequent current quench has a characteristic time scale of 30~ms, which is extended by an approximate factor of two by a loop voltage, which reaches 100~V, so the current quench is consistent with resistive dissipation.  
   
Figure 1 in a paper by de Vries et al \cite{de Vries:2016} is essentially a unique figure of a natural disruption in JET with an ITER-like wall.  The results are qualitatively different from those of Wesson et al.   The primary temperature collapse, from approximately 1~kev to 200~eV, and the current spike occur simultaneously, which means within the 1~ms time differences that can be distinguished on the published figure.  The internal inductance, which is a direct measure of the width of the current profile drops by a factor of two and remains low.  The decay time for the current is approximately 20~ms over the next 10~ms, which is consistent with a temperature of 17~eV, not 200~eV.  An obvious explanation would be that the current profile remains broad over that 10~ms due to magnetic field lines covering the plasma volume.  The destruction of magnetic helicity and hence the plasma current is then determined more by the edge than the central plasma temperature \cite{Boozer:pivotal}.  This evidence for the maintenance of chaotic magnetic field lines, rather than the fast re-formation of magnetic surfaces, is optimistic for the avoidance of runaways in the non-nuclear period of ITER operations, but the persistence of chaotic lines is unlikely to ensure the avoidance of runaway problems during nuclear operations on ITER \cite{Mitgation:Chalmers}.

The chaotic magnetic field lines produced in a fast magnetic reconnection will cause a rapid drop in the electron temperature---at least until the mean-free-path of the heat-carrying electrons becomes short compared to the distance required for a chaotic field lines to cross a large fraction of the plasma volume.  This is consistent with the data on DIII-D thermal quenches in Figure 10 of Paz-Soldan et al \cite{Paz-Soldan:2020}, which shows thermal quench times $\sim50~\mu$s.  Assuming a deuterium plasma and measuring electron density in $10^{20}$/m$^3$, the electron mean free path is $\lambda_e \approx 33 T^2/n$.  But, collisional heat transport along a magnetic field line is proportional the $T^{7/2}$, which implies the electrons that carry the heat have an energy of approximately $7T/2$.  Their mean free path, $\lambda_e^h$ is approximately twelve times longer than that of thermal electrons, $\lambda_e^h \approx 400 T^2/n$.  JET has a major radius of 3~m and a circumference of approximately 19~m, so the heat carrying electrons move through approximately $20 T^2/n$ toroidal transits between collisions.  The speed of the heat carrying electrons along the magnetic field lines is much faster than the Alfv\'en speed; the ratio is $v_e^h/V_A\approx 20 \sqrt{nT}/B$.  Electron cooling can also occur by radiation from impurities, and this is presumably required for a fast reduction of the electron temperature to values far below 1~keV.  The shortness of the electron mean-free-path at low temperatures prevents a rapid heat flow along chaotic magnetic field lines.  


\subsection{Drive and damping of Alfv\'en waves \label{sec:Alfven} }

As discussed in \cite{Boozer:acc}, a fast magnetic reconnection can be viewed as a quasi-ideal process, which conserves magnetic helicity and directly dissipates little energy.  Energy transfer out of the magnetic field  is given by $\vec{j}\cdot\vec{E}$.  In a fast magnetic reconnection, the dominant part is given by non-dissipative term, $\vec{u}\times\vec{B}$, in Ohm's law, $\vec{E}+\vec{u}\times\vec{B}=\vec{\mathcal{R}}$, namely $\vec{u}\cdot(\vec{j}\times\vec{B})$.  The condition $\vec{\nabla}\cdot\vec{j}=0$ implies that
\begin{eqnarray}
\vec{B}\cdot\vec{\nabla}\left(\frac{j_{||}}{B}\right)&=&\vec{B}\cdot\vec{\nabla}\times\left(\frac{\vec{f}_L}{B^2}\right) \nonumber\\
&=&\frac{\vec{B}\cdot\vec{\nabla}\times \vec{f}_L}{B^2}- \vec{B}\cdot\left(\vec{f}_L\times\vec{\nabla}\frac{1}{B^2}\right), \hspace{0.2in} \label{j_|| Lorentz} \\
\mbox{where   } && \vec{f}_L\equiv\vec{j}\times\vec{B}.
\end{eqnarray}
Any variation in $j_{||}/B$ along a magnetic field line implies a Lorentz force $\vec{f}_L$.  The first term on the right-hand side of Equation (\ref{j_|| Lorentz}) gives the variation in what is known as the the net $j_{||}/B$, which is zero along a magnetic field line in an equilibrium plasma, $\vec{f}_L=\vec{\nabla}p$, and the second term gives what is known as the Pfirsch-Schl\"uter variation in $j_{||}/B$. In a fast magnetic reconnection, two magnetic field lines with different magnitudes of $j_{||}/B$ can be quickly joined together, which makes  $\vec{B}\cdot\vec{\nabla}(j_{||}/B)=B\partial(j_{||}/B)/\partial\ell$ large and spatially complicated even in regions in which $\vec{\nabla}B^2$ is zero, where the Pfirsch-Schl\"uter term vanishes.  A curl of the Lorentz force is required.  In a scalar pressure, model of the plasma $\vec{f}_L=\rho \partial\vec{u}/\partial t +\vec{\nabla}p$.  Taking the density $\rho$ to be a spatial constant and letting $\hat{b}\equiv\vec{B}/B$, one finds that $\hat{b}\cdot\vec{\nabla}\times\vec{f}_L = \rho \partial\Omega/\partial t$, where $\Omega\equiv\hat{b}\cdot\vec{\nabla}\times\vec{u}$, the parallel component of the vorticity of the plasma flow.  As will be seen, this twisting motion drives a shear Alfv\'en wave.  

The propagation of Alfv\'en waves along chaotic field lines is thought to produce strong phase mixing and wave damping \cite{Heyvaerts-Priest:1983,Similon:1989}, which could heat the solar corona and slow the flattening of the $j_{||}/B$ profile.  But, the  flattening of the $j_{||}/B$ profile  appears to be approximately Alfv\'enic in tokamaks, and electron runaway provides a simpler explanation for corona formation, Appendix E of \cite{Boozer:acc}.  

On the sun, the footpoint motions of magnetic field lines naturally produce sufficiently large $j_{||}/B$'s, Appendix B of \cite{Boozer:ideal-ev}, for runaway with the short correlation distances across the field that are needed to avoid kinking.  The wave damping of \cite{Heyvaerts-Priest:1983,Similon:1989} is due to the exponentially increasing separation between neighboring chaotic lines.  But, the characteristic distance for an e-fold is apparently of order a thousand kilometers along magnetic field lines in the corona \cite{Boozer:acc}.  This is much longer than the height of the transition region above the photosphere, so exponentiation is unlikely to directly determine the height of the transition from the cold photospheric to the hot coronal plasma.


\section{Alfv\'en waves that flatten $K$ \label{sec:flattening} }

The standard assumptions of linearized reduced-MHD \cite{Kadomtsev,Strauss} will be made to derive the equations for the Alfv\'en waves that relax $\partial(j_{||}/B)/\partial\ell\rightarrow0$, where $\ell$ is the distance along a magnetic field line.  The required equations are simple and derived in \cite{Boozer:acc} and below for $K\equiv \mu_0j_{||}/B$ and for $\Omega\equiv \hat{b}\cdot\vec{\nabla}\times\vec{u}$, the vorticity along the magnetic field of the magnetic field line velocity:  
\begin{eqnarray}
&&\frac{\partial \Omega}{\partial t} = V_A^2  \frac{\partial K}{\partial \ell} + \nu_v \nabla_\bot^2 \Omega, \label{dOmega/dt} \\
&& \frac{\partial \Omega}{\partial\ell} = \frac{\partial K}{\partial t} -\frac{\eta}{\mu_0}\nabla^2_\bot K,  \label{dK/dt}
\end{eqnarray}
where $V_A$ is the Alfv\'en speed.  The field strength, plasma density $\rho$, the resistivity $\eta$, and the viscosity are assumed to be slowly varying in space and time when compared to $K$ and $\Omega$. The variables are time, the differential distance along a magnetic field line, $d\ell=R_0d\varphi$ in a torus, and  two coordinates across the field lines.  The spatial scale of the solution across the magnetic field lines, $\ell_\bot$, will be seen to be short compared to that along, $\ell_{||}$, so $\nabla^2\approx \nabla_\bot^2$.  

Although Equations (\ref{dOmega/dt}) and (\ref{dK/dt}) follow obviously from the linearized reduced-MHD equations, short derivations are sketched here for completeness.  Equation (\ref{dOmega/dt}) follows from the curl of the linearized force-balance equation, $\rho\partial\vec{u}/\partial t =-\vec{\nabla}p +\vec{j}\times\vec{B}-\rho\nu_v\nabla^2\vec{u}$.   The curl of the Lorentz force is $\vec{\nabla}\times(\vec{j}\times\vec{B})=\vec{B}\cdot\vec{\nabla}\vec{j} - \vec{j}\cdot\vec{\nabla}\vec{B}$.   The  component of $\vec{\nabla}\times(\vec{j}\times\vec{B})$ parallel to the magnetic field is approximated by $B \partial j_{||}/\partial \ell - j_{||} \partial B/\partial \ell  -  \vec{j}_\bot\cdot\vec{\nabla}B\approx  B^2 \partial(j_{||}/B)/\partial\ell$.  The current density $\vec{j}$ is divergence free, so $|j_{||}|/|\vec{j}_\bot|\sim \ell_{||}/\ell_\bot >>1$.  The gradients of the field strength across and along the magnetic field lines have more comparable scales.  The component of the curl of $\partial \vec{u}/\partial t$ that is parallel to the magnetic field gives Equation (\ref{dOmega/dt}).  

Equation  (\ref{dK/dt}), for the current evolution, follows from Ampere's law, Faraday's law, and Ohm's law, $\vec{E}+\vec{u}\times\vec{B}=\eta\vec{j}$.  The implication is $\mu_0\partial\vec{j}/\partial t =  \vec{\nabla}\times(\vec{\nabla}\times(\vec{u}\times\vec{B}-\eta\vec{j}))$.  A vector identity implies $\vec{\nabla}\times(\vec{u}\times\vec{B})=\vec{B}\cdot\vec{\nabla}\vec{u}-\vec{u}\cdot\vec{\nabla}\vec{B}$.  The parallel component of the $\partial\vec{j}/\partial t$ equation gives Equation  (\ref{dK/dt}).

The evolution equation for $K$ is obtained using the mixed-partials theorem applied to $\Omega$;
\begin{eqnarray}
 \frac{\partial^2K}{\partial t^2}- \frac{\partial }{\partial\ell}\left(V_A^2 \frac{\partial K}{\partial\ell}\right) = \left(\nu_v+\frac{\eta}{\mu_0}\right)\nabla_\bot^2 \frac{\partial K}{\partial t}. \
\end{eqnarray}
Neglecting the slow time dependence of the coefficients of the differential equation,
\begin{eqnarray}
 \omega^2 K + \frac{\partial }{\partial\ell}\left(V_A^2 \frac{\partial K}{\partial\ell}\right) = i\omega \left(\nu_v+\frac{\eta}{\mu_0}\right)\nabla_\bot^2K, \label{full wave eq}
\end{eqnarray}
where $\omega$ is a frequency.  The viscosity and resistivity are assumed to be small, so a term proportional to $\nu_v\eta$ has been ignored.  

Equation (\ref{full wave eq}) can be solved using the WKB method.  In this method, $K=K_s(\vec{x}_\bot) e^{iS}$, where the eikonal $S = S_0+S_1$ with $(\partial S/\partial t)_\ell = \omega$, so 
\begin{eqnarray}
&&\omega^2 K - V_A^2 \left(\frac{\partial S}{\partial\ell}\right)^2 K + K \frac{\partial}{\partial \ell}\left(iV_A^2 \frac{\partial S}{\partial\ell} \right) \nonumber\\
&& \hspace{1.0in} = i\omega \left(\nu_v+\frac{\eta}{\mu_0}\right)\nabla_\bot^2K
\end{eqnarray}
Choose $S_0$ so $(\partial S_0/\partial\ell)^2=\omega^2/V_A^2$.  The assumption is that the the parallel wavenumber, $k_{||}\equiv\partial S_0/\partial\ell$, varies slowly as a function of $\ell$, which would be exactly true if the coefficients in  Equation \ref{full wave eq} had no $\ell$ dependence.  There are two solutions: forward shear Alfv\'en waves moving in the direction of the field and backward waves moving in the opposite direction:
\begin{eqnarray}
S_{0f}&=& - \omega T_f \hspace{0.1in}\mbox{with} \hspace{0.1in}\\
T_f  &\equiv& t -\int\frac{d\ell}{V_A} \hspace{0.1in}\mbox{and} \hspace{0.1in}\\
S_{0b}&=& - \omega T_b \hspace{0.1in}\mbox{with} \hspace{0.1in}\\
T_b  &\equiv& t +\int\frac{d\ell}{V_A}.
\end{eqnarray}
The solution for the forward wave can be approximated $K=K_s e^{-i\omega T_f} e^{iS_1}$, where $S_1$ is slowly varying, and
\begin{eqnarray}
&& V_A^2 \left(2\frac{\partial S_{0f}}{\partial\ell}\frac{\partial S_{1f}}{\partial\ell}\right) K - K \frac{\partial}{\partial \ell}\left(iV_A^2 \frac{\partial S_{0f}}{\partial\ell} \right) \nonumber\\
&& \hspace{0.66in} = \left(\nu_v+\frac{\eta}{\mu_0}\right)\nabla_\bot^2\frac{\partial K}{\partial T_f}, \hspace{0.1in}\mbox{where} \hspace{0.2in} \\
&& \frac{\partial S_{0f}}{\partial\ell} =\frac{\omega}{V_A}.  \hspace{0.2in}\mbox{Consequently,} \\
&& 2 V_A \frac{\partial iS_{1f}}{\partial\ell} K' +  \frac{\partial V_A}{\partial \ell}K' \nonumber\\
&& \hspace{1.0in} = \left(\nu_v+\frac{\eta}{\mu_0}\right)\nabla_\bot^2K', \mbox{where} \hspace{0.3in}\\
&&K' \equiv \frac{\partial K}{\partial T_f} = - i\omega K = \left(\frac{\partial K}{\partial t}\right)_\ell.
\end{eqnarray} 
The resulting equation for the evolution of $K'$ is
\begin{eqnarray} 
&&\frac{1}{\sqrt{V_A}} \left(\frac{\partial \sqrt{V_A}K'}{\partial\ell} \right)_{T_f}= \frac{\Delta_d}{2}\nabla_\bot^2K'; \label{slow-ev}\\
&&\Delta_d \equiv \frac{\nu_v+\frac{\eta}{\mu_0}}{V_A} \\  \label{Delta-d}
&& \hspace{0.3in} \approx (1+P_{rm}) \frac{1.4\times10^{-8} n}{T^{3/2} B},
\end{eqnarray} 
where the magnetic Prandtl number $P_{rm}\equiv\mu_0\nu_v/\eta$.  $\Delta_d$ has units of length, meters, the number density has units of $10^{20}/$m$^3$, the temperature has units of keV, and the magnetic field has units of Tesla.  The solution for the backwards wave, which propagates in the negative $\ell$ direction, is identical except the sign of the righthand side.   

The cross-field ion viscosity $\nu_v$ is difficult to estimate, but the physics of the viscosity is closely related to that of the ion thermal transport.  If one assumes ion transport is gyro-Bohm-like then $\nu_v=(r_i/R_0) T/eB$ with $r_i$ the ion gyroradius and $R_0$ a typical spatial scale, such as the major radius of a tokamak.  Then, the magnetic Prandtl number is $P_{rm}\approx200 T^3/R_0B^2$.

The definition of $K'$ for a forward going wave can be understood.  Over  distances $\ell$ sufficiently short that $\sqrt{\Delta_d\ell}<<V_A/|\vec{\nabla}_\bot V_A|$, the functional form of $K$ is $K(t-\int\frac{d\ell}{V_A})$.  Letting a prime denote the derivative of $K(t-\int\frac{d\ell}{V_A})$ relative to its argument, $(\partial K/\partial t)_\ell= K'$ and $(\partial K/\partial \ell)_t = - K'/V_A$.

The interpretation of Equations  (\ref{full wave eq}) and (\ref{slow-ev}) is that shear Alfv\'en waves, which propagate along a magnetic field line with $d\ell/dt=\pm V_A$, serve as the basic characteristics for defining the solutions to Equation (\ref{full wave eq}) for $K$.  The part of $K$ that is not constant along the magnetic field line diffuses off the characteristics at the rate given by Equation (\ref{slow-ev}).   

When both $\Delta_d$ and $V_A$ are constant, the $K'$ in a magnetic flux tube obeys a conservation law---any change along the tube is due to diffusion through the sides.


\section{Monte-Carlo solution of Equation (\ref{slow-ev}) \label{sec:Monte-Carlo} }

\subsection{Initial $K'$}

Equation (\ref{slow-ev}) can be used to study the relaxation of $K'$ from an initial distribution $K'_0$.  The distribution of the parallel current, or more precisely the distribution of $K'$, along a magnetic field line immediately after magnetic surfaces have broken can be calculated using the dominance of the dependence of $K_0$ on $T_f$.  Since $\vec{B}\cdot\vec{\nabla}K_0= K'_0 \vec{B}\cdot\vec{\nabla}T_f=-(B/V_A)K'_0$,
\begin{equation}
K'_0=-V_A\frac{\vec{B}\cdot\vec{\nabla}K_0}{B}. \label{K'_0}
\end{equation}
for the forward wave.  The sign of the righthand side is opposite for the backwards wave.  For the forward wave, $K'$ propagates along the magnetic field lines at the Alfv\'en speed, $d\ell/dt = V_A$, and diffuses off the lines at the slow rate given by Equation (\ref{slow-ev}). 

\subsection{Monte Carlo operator}

Equation (\ref{slow-ev}) can be solved using a Monte Carlo approach that is derived in Section IV of \cite{Boozer:Monte Carlo}.   The term $\nabla^2_\bot K'$ can be calculated using ordinary $R,Z$ cylindrical coordinates for a tokamak since the toroidal magnetic field is assumed far stronger than the poloidal.  In the large aspect ratio limit
\begin{equation}
\nabla_\bot^2 K'=\frac{\partial^2K'}{\partial R^2} + \frac{\partial^2K'}{\partial Z^2},
\end{equation}
where $R$ and $Z$ are the position of a particular magnetic field line as it is followed using the distance along the line $\ell=R_0\varphi$.  

Equation (\ref{slow-ev}) implies that when $K'$ is non-zero only within a small range of $R$ and $Z$ then at a constant $T_f$ the function $K'(\ell,R,Z,T_f)$ obeys
\begin{eqnarray}
&&\frac{\partial  \int R K'dRdZ}{\partial \ell} =  \nonumber\\
&& \hspace{0.2in}\frac{\Delta_d}{2} \int R \left(\frac{\partial^2K'}{\partial R^2} + \frac{\partial^2K'}{\partial Z^2}\right)dRdZ=  \nonumber\\
&& \hspace{0.2in}\frac{\Delta_d}{2} \int \frac{\partial}{\partial R} \left(R\frac{\partial K'}{\partial R} - K' \right)dRdZ =0. \hspace{0.3in} 
\end{eqnarray}
This equation and the similar equation for $\int ZK'dRdZ$ imply there is no systematic drift of $K'$ off the field line.  But, $K'$ does diffuse off of the field line for
\begin{eqnarray}
&&\frac{\partial  \int R^2K'dRdZ}{\partial \ell} =  \nonumber\\ 
&& \hspace{0.2in} \frac{\Delta_d}{2} \int R^2\left(\frac{\partial^2K'}{\partial R^2} + \frac{\partial^2K'}{\partial Z^2}\right)dRdZ =\nonumber \\
&&\hspace{1.0in} \Delta_d \int K'dRdZ
\end{eqnarray}
with a similar equation for $\int Z^2K'dRdZ$.  Following the Monte-Carlo derivation in Section IV of \cite{Boozer:Monte Carlo}, the interpretation is that when $K'$ is a delta function about $R_s,Z_s$ before the application of Equation (\ref{slow-ev}), then after the application, $K'$ will have a Gaussian distribution about the point $R_s,Z_s$ with a standard deviation given by $\partial\sigma^2/\partial\ell=\Delta_d$.   

Each small step $\delta\ell$ along a magnetic field line consists of two operations: (1) The $R$ and $Z$ are changed to track a particular line.  (2)  Steps $\delta R=\pm\sqrt{ \Delta_d\delta\ell}$ and $\delta Z=\pm \sqrt{ \Delta_d\delta\ell}$ are taken to a new field line.  The integration can then be followed for another $\delta\ell$ step.  The symbol $\pm$ implies the sign is chosen with equal probability of being plus or minus. The advance in time during a step is $\delta t= \delta\ell/V_A$ for the forward moving and $\delta t= -\delta\ell/V_A$ for the backward moving wave.

\subsection{A study of the flattening} 

The chaotic magnetic field that arises in a disruption simulation can be used to study flattening of the current profile.  To do this the plasma volume can be separated into cells, each with the same volume.  The initial $K'_0$ can be obtained by superimposing the parallel current distribution in the pre-disruption plasma on the chaotic magnetic field and using Equation (\ref{K'_0}) to find a value for $K'_0$ in each cell.   Start $N_0$ trajectories in each cell with half propagating forward and half propagating backward along the field lines.  The value of $K'_j(t)$ in cell $j$ at time $t$ is the sum of the $K'_i(0)$ that are now in cell $j$, starting in cell $i$ at $t=0$ divided by $N_0$.  The statistical error scales as $1/\sqrt{N_0}$.

The magnetic field lines and the volume in which they are chaotic change over the time scale of the current flattening.  This can be studied by updating the field line trajectories as the current profile flattens.  Before each step, $\delta t=\pm\delta\ell/V_A$, the magnetic field line trajectories should be updated, and $K'_0$ in each cell at the beginning of the new step is given by Equation (\ref{K'_0}).  This should be calculated using the part of the parallel current that is independent of the non-inertial forces, such as the pressure gradient.  The part of the parallel current driven by non-inertial forces, such as the pressure gradient, is the Pfirsch-Schl\"uter current.

\subsection{Alfv\'en wave reflection}

In a tokamak, the wall is not normally a magnetic surface; it is penetrated by what is known as the vertical magnetic field.  An implication is that a region of chaotic magnetic field lines can extend all the way to the walls.  The Alfv\'en waves that give the relaxation of $K'$ are naturally reflected by the walls---either by perfectly insulating or by perfectly conducting walls---but the sign of the reflected wave is opposite in the two cases.  Wave reflection switches the characteristic that the wave is following.  When the WKB approximation is valid, which requires $k_{||}$ change slowly with respect to $\ell$, a switch in the wave from following one characteristic to the  other is not possible.

\subsubsection{Reflection from an insulating wall}

When the wall is a perfect insulator, $K=0$ on the wall.  A steady state current cannot flow along a chaotic field line that strikes an insulating wall, and the reflected Alfv\'en waves serve to cancel $K'$.  The net parallel current  drops to zero in an outer region of chaotic field lines on the time scale for a shear Alfv\'en wave to traverse the region by propagating along the chaotic field lines.

The decay of the current after the current spike appears to be far slower than the time it takes an Alfv\'en wave propagating along chaotic field lines to reach the walls, which implies the insulating-wall boundary condition $K=0$ is not  realistic.   The flux of magnetic helicity along the magnetic field lines, which is denoted by $2\mathcal{F}_{||}$ in  \cite{{Boozer:runaway-ITER}} is not blocked by a wall that is a perfect insulator but is when the wall is a perfect conductor.

\subsubsection{Reflection by drag}

Even a perfectly conducting medium can exert a drag force on the motion of the magnetic field lines, which is balanced by the Lorentz force that causes a change in $K=\mu_0 j_{||}/B$, Equation (\ref{j_|| Lorentz}).  

The drag force can be quantified by a drag time $\tau_d$.  In one dimension plus time, the equations are
\begin{eqnarray}
 \frac{\partial\Omega}{\partial t} = V_A^2\frac{\partial K}{\partial \ell} -\frac{\Omega}{\tau_d(\ell)} \hspace{0.2in}\mbox{and}\hspace{0.2in} \frac{\partial\Omega}{\partial\ell} = \frac{\partial K}{\partial t}.  \label{d Omega / d ell}
 \end{eqnarray}
 The mixed-partials theorem applied to $K$ implies
 \begin{equation}
 V_A^2  \frac{\partial^2\Omega}{\partial\ell^2}= \frac{\partial^2\Omega}{\partial t^2} +\frac{1}{\tau_d} \frac{\partial\Omega}{\partial t}. \label{drag eq}
 \end{equation}
 The drag, which is proportional to $1/\tau_d$, will be assumed to be zero for $\ell<\ell_0$ but a non-zero constant for $\ell>\ell_0$.  The wave equation for $\Omega$ is simpler than the equation for $K$ since that equation includes a term proportional to $d(1/\tau)/d\ell$.  In the two regions in which $\tau_d$ is constant, Equation (\ref{drag eq}) can be solved by $\Omega \propto \exp\big(i(k\ell -\omega t)\big)$.  Let
 \begin{eqnarray}
 k_A &\equiv& \frac{\omega}{V_A} \hspace{0.2in}\mbox{and}\hspace{0.2in} \ell_d \equiv V_A\tau_d,  \hspace{0.2in}\mbox{then}\hspace{0.2in} \\
 k_\pm&=&\pm k_A \sqrt{1+\frac{ i}{\Lambda_d}}, \hspace{0.2in}\mbox{where}\hspace{0.2in} \Lambda_d\equiv k_A\ell_d.\\
\Omega&=&\mathcal{R}_\Omega e^{i(k_+\ell-\omega t)} \hspace{0.2in}\mbox{for}\hspace{0.1in} \ell>\ell_0 \\
&=&\left(R_\Omega e^{ik_A\ell} + L_\Omega e^{-ik_A\ell}\right) e ^{-i\omega t} \hspace{0.1in}\mbox{for}\hspace{0.1in} \ell<\ell_0   \hspace{0.2in}. 
\end{eqnarray}
Neither $\Omega$ nor $\partial\Omega/\partial\ell$ is discontinuous at $\ell_0$, so $\mathcal{R}_\Omega= R_\Omega+ L_\Omega$ and $k_+\mathcal{R}_\Omega=k_A(L_\Omega-R_\Omega)$, which imply
\begin{eqnarray}
L_\Omega&=& -\frac{\sqrt{1+\frac{i}{\Lambda_d}}-1}{\sqrt{1+\frac{i}{\Lambda_d}}+1}R_\Omega; \\
\mathcal{R}_\Omega &=& \frac{2}{\sqrt{1+\frac{i}{\Lambda_d}}+1}R_\Omega.
\end{eqnarray}
Equation (\ref{d Omega / d ell}) implies $K=(i/\omega)\partial\Omega/\partial\ell$ has the same form as $\Omega$ but with coefficients  $\mathcal{R}_K$, $R_K$, and $L_K$.
\begin{eqnarray}
\mathcal{R}_K &=& -\frac{2\sqrt{1+\frac{i}{\Lambda_d} } }{\sqrt{1+\frac{i}{\Lambda_d}}+1}\frac{R_\Omega}{V_A};\\ \nonumber\\
R_K &=&-\frac{R_\Omega}{V_A};\\
L_K &=& - \frac{\sqrt{1+\frac{i}{\Lambda_d}}-1}{\sqrt{1+\frac{i}{\Lambda_d}}+1}\frac{R_\Omega}{V_A};\\
R_K+L_K&=&-\frac{2\sqrt{1+\frac{i}{\Lambda_d}} }{\sqrt{1+\frac{i}{\Lambda_d}}+1}\frac{R_\Omega}{V_A}= \mathcal{R}_K,
\end{eqnarray}
Both the vorticity $\Omega$ and the parallel current or $K$ are continuous at $\ell_0$, the location at which the drag jumps from zero to a finite value.  A strong drag, $\Lambda_d\rightarrow0$, implies the wave is stopped in a far shorter distance than a wavelength, which reflects the wave perfectly.  When $R_K$ is the amplitude of the parallel current function propagating towards the region of strong damping, $L_K=R_K$ is the amplitude of  the reflected wave propagating away.  When small but non-zero $\Lambda_d$ effects are retained, $L_K/R_k = 1+(i-1)\sqrt{2\Lambda_d}$.  The imaginary term is equivalent to a time delay.

\subsubsection{Reflection by a jump in Alfv\'en speed}

A sudden change in the Alfv\'en speed at $\ell=\ell_0$ will also violate the WKB approximation.  Assume the Alfv\'en speed jumps from $V_n$ to $V_p$ as $z\equiv\ell-\ell_0$ goes from negative to positive.  This boundary condition is probably not applicable to a tokamak with chaotic field lines at its edge but is of interest for solar problems. 

The Alfv\'en equation for the parallel current $K\equiv \mu_0j_{||}/B$ is
\begin{equation}
\frac{\partial^2K}{\partial t^2} =  \frac{\partial}{\partial z}\left(V_A^2(z) \frac{\partial K}{\partial z}\right).
\end{equation}
The solution for a wave launched so it is going to the right from the negative $z$ side is
\begin{eqnarray}
K &= R_n e^{i(k_nz-\omega t)} + L_n e^{-i(k_nz+\omega t)} \hspace{0.1in} &z<0;\hspace{0.2in}\\
&= R_p e^{i(k_nz-\omega t)}  &z>0,
\end{eqnarray}
where $k_n = \omega/V_n$ and $k_p=\omega/V_p$.

Two conditions must be satisfied at $z=0$, the continuity of $K$ and the continuity of $V_A^2 \partial K/\partial z$.  These two conditions imply
\begin{eqnarray}
R_n + L_n &=& R_p; \\
V_n^2 k_n(R_n-L_n) &=& V_p^2 k_pR_p, \hspace{0.1in}\mbox{or} \hspace{0.2in}\\
V_n(R_n-L_n) & =& V_p R_p.
\end{eqnarray}
The solution is
\begin{eqnarray}
L_n &=& \frac{V_n-V_p}{V_p+V_n}R_n;\\
R_p &=& \frac{2V_n}{V_p+V_n}R_n.
\end{eqnarray}

When an Alfv\'en wave carrying $K$ propagates from the solar photosphere towards the corona, the Alfv\'en speed undergoes a large increase, which implies $K$ is reduced in amplitude by a factor $V_n/V_p$ on the corona side from the incoming $K$ on the photosphere side.  In the limit as $V_n/V_p\rightarrow0$, the boundary acts as an insulator when viewed from the photosphere.  

An Alfv\'en wave propagating from the corona towards the photosphere undergoes a large reduction in the Alfv\'en speed, which when over a sufficiently short spatial scale, causes a reflection of the wave back into the corona but with the amplitude of $K$ in the photosphere having twice the amplitude as in the incoming $K$ in the corona. 


\section{Discussion \label{sec:discussion} }

Understanding the physical states through which ITER may evolve during disruptions is essential for an assessment of how the issue of runaway electrons can be managed to minimize the impact on the ITER mission.  Much of this data is encoded in the flattening of $K\equiv\mu_0j_{||}/B$, and this defines the importance of derivations given in this paper.

As has been known for almost thirty years  \cite{Jardin:1991}, the rapid breaking of magnetic surfaces and helicity conservation are fundamental to the physics of current spikes.  For a current spike to be observed, the time scale for the flattening of the parallel current density $j_{||}$ must be short in comparison to the resistive dissipation of the current; the reconnection must be fast.  Current spikes and magnetic reconnections have been seen in three-dimensional NIMROD simulations in 2010 by Izzo and Parks \cite{Izzo:2010}.  Eric Nardon and collaborators have made related calculations with the JOREK code \cite{Nardon:JOREK2017}.

Three-dimensional simulations of large tokamaks, but especially ITER, are computationally demanding, so only a few cases can be studied, and even these contain simplifying assumptions.  Their reliability and utility depend on understanding the physical and mathematical reasons for the results.  From the mathematics of fast magnetic reconnection, one expects flux tubes in annular regions of intact magnetic surfaces to show exponentially large distortions in the cross-sectional shape as that annular region evolves toward a state in which fast magnetic reconnection occurs.  Reconnection occurs when resistive diffusion across the thinest part of a flux tube can compete with the evolution time scale.  Unfortunately, no one has documented this effect in tokamak disruption simulations, but Daughton et al \cite{Daughton:2014} studied the relation between reconnection regions and large exponential separations of of neighboring magnetic field lines and found a close relation.

There are two parts to the rapid flattening of the current: (1) a fast magnetic reconnection of the surfaces, which conserves magnetic helicity \cite{Boozer:ideal-ev,Boozer:surf-loss,Boozer:acc,Boozer:pivotal} and (2) a flattening of the parallel current along the newly chaotic magnetic field lines by Alfv\'en waves \cite{Boozer:acc}.  Alfv\'en waves propagating along chaotic field lines can be heavily damped \cite{Heyvaerts-Priest:1983,Similon:1989}, which could in principle extend the time required for the flattening sufficiently to eliminate current spikes.  This paper found the viscosity $\nu_v$ and the resistivity $\eta$ diffusively spread shear Alfv\'en waves across the field lines by a distance $\approx\sqrt{\Delta_d\ell_p}$, where the distance $\Delta_d$ is given in Equation (\ref{Delta-d}) and $\ell_p\approx 100\times 2\pi R_0$ is the distance Alfv\'en waves must propagate along the field lines to flatten the current.  Using the estimates for $\Delta_d$ and $\ell_p$ given in the paper, the distance $\sqrt{\Delta_d\ell_p}$ appears to be of order centimeters, which seems unlikely to significantly slow the flattening.  The Monte Carlo methods developed in the paper together with numerical models of the chaotic magnetic fields of a tokamak disruption could be used to determine how large $\Delta_d$ would have to be to significantly slow the flattening of the current.

\vspace{0.2in}

\section*{Acknowledgements}

This material is based upon work supported by the U.S. Department of Energy, Office of Science, Office of Fusion Energy Sciences under Award Numbers DE-FG02-03ER54696, DE-SC0018424, and DE-SC0019479.   

\section*{Data availability statement}

Data sharing is not applicable to this article as no new data were created or analyzed in this study.



\end{document}